\documentclass[aps,prb,twocolumn,groupedaddress,showpacs]{revtex4-1}
\usepackage{graphics}
\usepackage{amssymb}
\usepackage{graphicx}
\usepackage[reqno]{amsmath}
\usepackage{epsfig}
\usepackage{amsmath}

\DeclareGraphicsExtensions{.pdf,.eps,.png,.jpg,.mps}

\begin{document}
\title{Bound state in the continuum and spin filter in artificial molecules}
\author{J. P. Ramos}
\affiliation{Departamento de F\'{i}sica, Universidad Cat\'{o}lica del Norte,
Casilla 1280, Antofagasta, Chile}
\author{P. A. Orellana}
\affiliation{Departamento de F\'isica, Universidad T\'ecnica Federico Santa Mar\'ia, Casilla 110 V, Valpara\'iso, Chile}
\date{\today}

\begin{abstract}

In this paper we study the formation of bound states in the continuum in a quantum dot molecule coupled to leads. Based in the combination of bound states in the continuum and Fano effect, we propose a new design of a spin-dependent polarizer. By lifting the spin degeneracy of the carriers in the quantum dots by means of a magnetic field the system can be used as a  spin-polarized device. A detailed analysis of the spin-dependent conductance and spin polarization as a function of the applied magnetic field and gate voltages is carried out.

\end{abstract}
\maketitle
\section{Introduction}\label{intro}

Spintronics is a wide field of research that exploits the spin degree of freedom of the electrons \cite{pulizzi}. In spintronics, the precise control of spin currents leads to new functionalities and higher efficiencies than conventional electronics. In the last years, there have been a renewed interest in the design and use of nanostructures as spintronic devices in which the spin degree of freedom is used as the basis for their operation \cite{Intro1, Intro2, song, Intro3, Intro4}. These devices exploit the natural connection between the conductance and the quantum-mechanical transmission, in order to generate and detect spin polarized currents.

The quantum dots are suitable systems to be used as spintronic devices. They are artificial nanostructures in which electrons can be confined even in all space dimensions~\cite{QDs, QD}, resulting in charge and energy quantization. Both features present in real atomic systems, so useful analogies between real and artificial atomic systems have been exploited recently. By enforcing this analogy, effects like Fano~\cite{fano} and Dicke~\cite{dicke} have also been found to be present in quantum dot configurations. In this context, Song et al.~\cite{song} described how a spin filter may be achieved in open quantum dot systems by exploiting the Fano resonances occurring in their transmission characteristics. In a quantum dot in which the spin degeneracy of a carrier has been lifted, they showed that the Fano effect might be used as an effective means to generate spin polarization of transmitted carriers. The electrical detection of the resulting spin polarization should then be possible.

In a previous work~\cite{OrAd}, we have shown that in a side-coupled double quantum dot system the transmission shows a large peak-to-valley ratio. The difference in energy between the resonances and antiresonances can be controlled by adjusting the difference between the energy levels of the two quantum dots by gate voltages. The above behavior in the transmission is a result of the quantum interference of electrons through the two different resonant states of the quantum dots, coupled to common leads. This phenomenon is analogous to the Dicke effect in quantum optics which takes place in the spontaneous emission of two closely lying atoms radiating a photon into the same environment. In the electronic case, however, the decay rates (level broadening) are produced by the indirect coupling of the up-down QDs, giving rise to a fast (superradiant) and a slow (subradiant) mode. Also,  M. L. Vallejo et al.\cite{marcela} have shown that the coexistence of a bound state in continuum (BIC) and Fano antiresonance can be exploited to result in polarizations close to $100\%$ in wide regions in the space of used parameters. 

The existence of BICs was predicted at the dawn of quantum mechanics by von Neumann and Wigner \cite{vonneu} for certain spatially oscillating attractive potentials, for a one-particle  Sch\"odinger  equation. Much later, Stillinger and Herrick \cite{Stillinger} generalized von Neumann's work by analyzing a two-electron problem, they found BICs were formed despite the interaction between electrons. BICs have also shown to be present in the electronic transport in meso and nanostructures \cite{Schult,Nockel,rotter,loreto,jhon}.  Several mechanisms of formation of BICs in open quantum dots (QDs) have been reported in the literature. The simplest one is based on the symmetry of the systems and, as a consequence, in the difference of parity between the QD eingenstates and the continuum spectrum \cite{Texier}.

In the present work, we study the formation of BICs in a system formed by a quantum dot coupled to two leads with a side attached molecule. We show that the combination of BICs and Fano effect can be used to design a spin filter. By tuning the positions of BICs and Fano antiresonances with a magnetic field and gate voltages, this system can be used as an efficient spin filter even for small values of magnetic fields.

\section{Model}\label{model}

\begin{figure}[ht]
\begin{center}
\includegraphics[width=8cm]{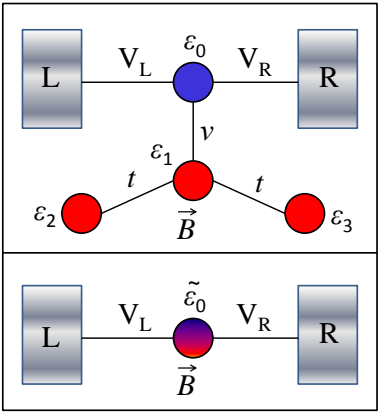}\\
  \caption{(Color online) Upper panel: Quantum dot  coupled to two leads with a side attached molecule . Lower panel: Effective model, derived by a decimation process.}\label{fig1}
\end{center}
\end{figure}

The system  is formed by a quantum dot  coupled to two leads with a side attached molecule, as is shown schematically in Fig.~(\ref{fig1}, upper panel).

The full system is modeled by a generalized non-interacting Anderson Hamiltonian, namely  $H=H_{\text{l}}+H_{\text{m}}+H_{\text{in}}$, where $H_{\text{l}}$ represents the Hamiltonian of the leads, $H_{\text{m}}$ the Hamiltonian of the embedded quantum dot with a side attached quantum dot molecule and $H_{\text{in}}$ represents the interaction between the leads and the embedded quantum dot. We consider each quantum dot with a single energy level $\varepsilon_{i,\sigma}$ ($i=1,..4$). Then,
\begin{equation}\label{eq1}
    H_{\text{l}}=\sum_{\sigma\,;\,\alpha =L,R}\varepsilon_{k_{\alpha}}c_{k_{\alpha},\sigma}^{\dag}c_{k_{\alpha},\sigma}\,,
\end{equation}
\begin{eqnarray}\label{eq2}
     H_{\text{m}} &=&\sum_{i=0,\sigma}^{3}\varepsilon_{i,\sigma}d_{i,\sigma}^{\dag}d_{i,\sigma}-t\sum_{j=2,3,\sigma}(d_{1,\sigma}^{\dag}d_{j,\sigma}+\text{h.c.})\\ \nonumber
    &-&\sum_{\sigma}\nu(d_{0,\sigma}^{\dag}d_{1,\sigma}+\text{h.c.})\,,
\end{eqnarray}
\begin{equation}\label{eq3}
    H_{\text{in}}=\sum_{\sigma\,;\,\alpha =L,R}(V_{\alpha} d_{0,\sigma}^{\dag}c_{k_{\alpha},\sigma}+\text{h.c.})\, ,
\end{equation}
where $d_{i,\sigma}^{\dag}(d_{i,\sigma})$ is the electron creation (annihilation) operator in the i-th QD with spin $\sigma$ ($\sigma=\downarrow,\uparrow$), the $c_{k_{\alpha}}^{\dag}(c_{k_{\alpha}})$ is the electron creation (annihilation) operator in the lead $\alpha$ ($\alpha = L,R$) with momentum $k$ and spin $\sigma$; while $V_{\alpha}$ is the tunneling coupling between the $\alpha$ lead and the embedded quantum dot.  Besides, $\nu$ is the coupling between the embedded quantum dot and the quantum dot molecule and $t$ the interdot coupling in the side attached quantum dot molecule. We consider a system in equilibrium, at zero temperature with $\varepsilon_{i,\sigma}=\varepsilon_{i}+\sigma \varepsilon_z$ ($i=0,1,2,3$) where $\varepsilon_{i}$ is the energy level of the isolated $i$ quantum dot, $\varepsilon_{z}= g\mu_B B$ is the Zeeman energy due to the presence of the magnetic field $B$, with $g$  the electron Land\'e factor and $\mu_B$ the Bohr magneton.

By using a re-normalization process based in the the Dyson equation for the Green's functions (see the appendix), we obtained an effective model for the artificial molecule. Fig.\ref{fig1} (lower panel)  displays the effective model which consists in a single quantum dot coupled to two lead with an effective energy $\tilde{\varepsilon}_{0,\sigma}$,
\begin{equation}\label{eq4}
\tilde{\varepsilon}_{0,\sigma} = \varepsilon_{0,\sigma}+\Sigma_{\text{m},\sigma}(\varepsilon)\,,
\end{equation}
where $\varepsilon_{0,\sigma} \equiv \varepsilon_{0} + \varepsilon_{z}\sigma_{\sigma\bar{\sigma}}^{z}$, and $\Sigma_{\text{m},\sigma}(\varepsilon)$ is the self-energy due the presence of the side attached molecule given by
\begin{equation}\label{eq5}
\Sigma_{\text{m},\sigma}(\varepsilon)=\frac{\nu^{2}}{\varepsilon-\varepsilon_{1,\sigma}-\frac{t^2}{\varepsilon-\varepsilon_{2,\sigma}}-\frac{t^2}{\varepsilon-\varepsilon_{3,\sigma}}}\,.
\end{equation}

By using the Dyson's equation $G=g_{0}+g_{0}H_{\text{in}}G$, where $G$ is the Green's function of the system and $g_{0}$ the Green's function of the effective quantum dot, the Green's function of the effective quantum dot coupled to two lead is obtained
\begin{equation}
\label{eq6}
G_{0,\sigma}(\varepsilon)=\frac{1}{\varepsilon-\tilde{\varepsilon}_{0,\sigma}-\Sigma_{\text{l}}(\varepsilon)}\,,
\end{equation}
where $\Sigma_{\text{l}}(\varepsilon)$ is the self energy due to leads.

%$\Gamma=-i\Sigma_L$ describing the coupling between contact--molecule.%

Once we obtain the Green's function $G_{0,\sigma}$, we can evaluate the spin dependent transmission probability $\tau_{\sigma}$ by using the Fisher-Lee relation~\cite{fisherlee}
\begin{equation}\label{eq7}
\tau_{\sigma}(\varepsilon) =\frac{\Gamma^2}{(\varepsilon-\tilde{\varepsilon}_{0,\sigma})^2+\Gamma^2}\,,
\end{equation}
where $\Gamma=-i\Sigma_{\text{l}}$, is the effective coupling between the effective quantum dot with the leads.

The linear conductance $\mathcal{G}$ is obtained by the Landauer's formula at zero temperature, $\mathcal{G}_{\sigma}(\varepsilon_{f})=(e^2/h)\tau_{\sigma}(\varepsilon=\varepsilon_{f})$~\cite{Landauer}. Then, the linear conductance of the system is given by
\begin{equation}\label{eq8}
    \mathcal{G}_{\sigma}(\varepsilon_{f})=\frac{e^2}{h}\frac{\Gamma^2}{(\varepsilon_{f}-\varepsilon_{0,\sigma}-\Sigma_{\text{m}}(\varepsilon_{f}))^2+\Gamma^2}\,.
\end{equation}

The resonances in the linear conductance are obtained by the equation
\begin{equation}\label{eq9}
\varepsilon_{f}-\varepsilon_{0,\sigma}-\Sigma_{\text{m},\sigma}(\varepsilon_{f})=0\,,
\end{equation}
while the zeros of the conductance coincide with the poles of the self-energy $\Sigma_{\text{m},\sigma}(\varepsilon_{f})$, which correspond to the spectrum of the side attached molecule.

For simplicity, we setting the energy levels of the side attached molecule as, $\varepsilon_{1,\sigma}=\varepsilon_{0,\sigma}$, $\varepsilon_{2,\sigma}=\varepsilon_{0,\sigma}+\Delta$, $\varepsilon_{3,\sigma}=\varepsilon_{0,\sigma}-\Delta$. On one hand, the resonances are obtained from the eigenvalues of Eq.~(\ref{eq2})
\begin{subequations}
\begin{eqnarray}\label{eq10}
      2(\varepsilon_{\pm,\sigma}^{+}-\varepsilon_{0,\sigma})^2&=&\Delta^{2}+\nu^{2}+2t^{2}-\sqrt{\xi}\,,\\
      2(\varepsilon_{\pm,\sigma}^{-}-\varepsilon_{0,\sigma})^2&=&\Delta^{2}+\nu^{2}+2t^{2}+\sqrt{\xi}\,,
\end{eqnarray}
\end{subequations}
where $\xi=(\Delta^{2}-\nu^{2})^{2}+4t^{2}(\Delta^{2}+\nu^{2})+4t^{4}$. On the other hand, the antiresonances are obtained from the spectrum of the side attached molecule, which it is given by,
\begin{subequations}
\begin{eqnarray}\label{eq11}
    \omega_{1,\sigma} &=&\varepsilon_{0,\sigma}\,, \\
    \omega_{2,\sigma} &=&\varepsilon_{0,\sigma}+\sqrt{2t^2+\Delta^2}\,, \\
    \omega_{3,\sigma} &=&\varepsilon_{0,\sigma}-\sqrt{2t^2+\Delta^2}\,.
\end{eqnarray}
\end{subequations}

We can rewritten the self-energy $\Sigma_{\text{m},\sigma}(\varepsilon)$ in terms of the spectrum of the side attached molecule as following,

\begin{subequations}
\begin{eqnarray}
\label{eq12}
\Sigma_{\text{m},\sigma}(\varepsilon)&=&\Sigma_{\text{m}1,\sigma}(\varepsilon)+\Sigma_{\text{m}2,\sigma}(\varepsilon)+\Sigma_{\text{m}3,\sigma}(\varepsilon)\,,\\
\Sigma_{\text{m}\alpha,\sigma}(\varepsilon)&=&\frac{\nu_{\alpha}^{2}}{\varepsilon-\omega_{\alpha,\sigma}},\,\,\,\,\,(\alpha=1,2,3)\,,\\
\nu_{1}^{2}&=&\frac{\Delta^2\nu^2}{\Delta^2+2t^2}\,,\\
\nu_{2}^{2}&=&\nu_{3}^{2}=\frac{t^2\nu^2}{\Delta^2+2t^2}\,.
\end{eqnarray}
\end{subequations}

In order to quantify the degree of spin polarization, we introduce the weighted spin polarization as~\cite{song}
\begin{equation}\label{eq13}
P_{\sigma}=\frac{\vert \tau_{\uparrow}-\tau_{\downarrow}\vert}{\vert \tau_{\uparrow}+\tau_{\downarrow}\vert } \tau_{\sigma}\,.
\end{equation}
Notice that this definition takes into account not only the relative fraction of one of the spins, but also the contribution
of those spins to the electric current. In other words, we will require that not only the first term of the right-hand side of Eq.~(\ref{eq13}) to be of the order of unity, but also the transmission probability $\tau_{\sigma}$.

\section{Results}
\label{resultados}
In what follows all energies will be given in units of the parameter $\Gamma$ and we setting $t=\nu=\Gamma/2$ and $\varepsilon_0=0$.

\begin{figure}[h!]
\begin{center}
  \includegraphics[width=7cm]{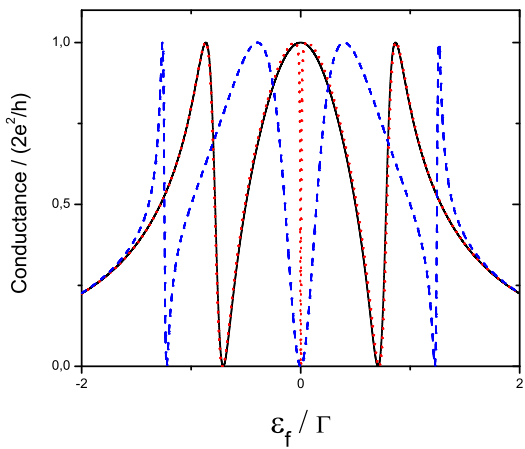}\\
  \caption{(Color online) Linear conductance as a function of the Fermi energy  for $\Delta=0$ (solid black line), $\Delta=0.1\,\Gamma$ (dotted red line) and $\Delta=\Gamma$ (dashed blue line), in absence of magnetic field.}
 \label{fig2}
\end{center}
\end{figure}

\begin{figure}[h!]
\begin{center}
  \includegraphics[width=7cm]{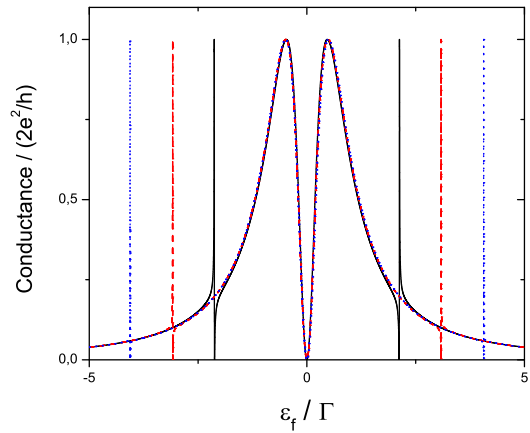}\\
  \caption{(Color online) Linear conductance as a function of the Fermi energy  for $\Delta=2\Gamma$ (solid black line),  $\Delta=3\,\Gamma$ (dotted red line) and $\Delta=4\Gamma$ (dashed blue line), in absence of magnetic field.}
  \label{fig3}
\end{center}
\end{figure}

\begin{figure}[h!]
\begin{center}
\includegraphics[width=9cm]{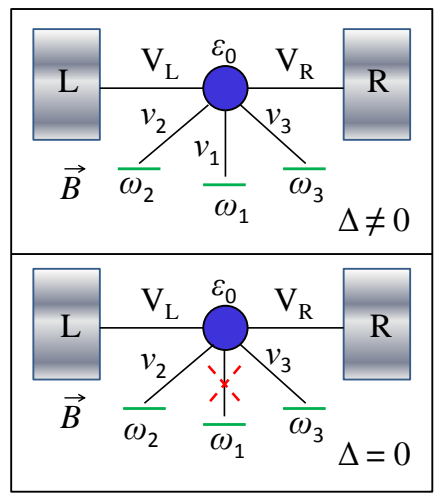}\\
\caption{(Color online) Schematic view of the molecular states with the embedded quantum dot, for $\Delta \neq 0$, (upper panel) and $\Delta=0$ (lower panel).}\label{fig4}
\end{center}
\end{figure}

In first place, we show results for the linear conductance in absence of magnetic field. The Fig (\ref{fig2}) displays the linear conductance versus the Fermi energy  for $\Delta=0$ (solid black line), $\Delta=0.1\Gamma$ (dotted red line) and $\Delta=\Gamma$ (dashed blue line). For $\Delta=0$ the conductance shows two Fano antiresonances and three resonances. The Fano antiresonances arise from the destructive interference between the continuum state of the embedded quantum dot and the localized states in the side-attached molecule \citep{kobayashi}.
From Eqs.~(\ref{eq12}-c) we can see that when $\Delta = 0$, the molecular state with energy $\omega_{1,\sigma}=\varepsilon_{0,\sigma}$ decouples from the continuum and therefore does not contribute to the conductance ($\Sigma_{\text{m}1,\sigma}(\varepsilon)=0$, $\nu_1=0$). This state becomes a BIC. This state  arises from the hybridization of the states of the quantum dots 2 and 3 through the quantum dot 1 and has no projection on the continuum, as it is shown schematically in Fig.~(\ref{fig4}). Besides, we see from Eq.~(\ref{eq10}) that two resonances collapse at $\varepsilon_f=0$.

\begin{figure}[h]
\begin{center}
\includegraphics[width=7cm]{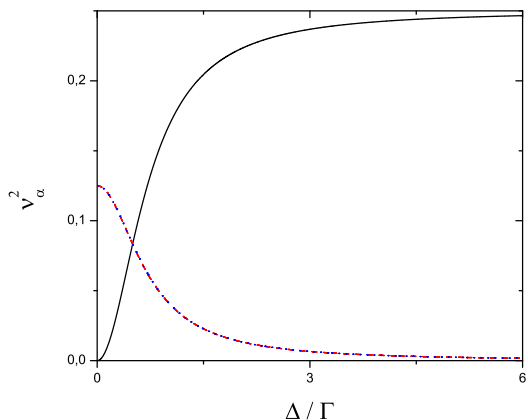}\\
  \caption{(Color online) Square effective coupling ($\nu_{\alpha}^2$) as a function of $\Delta$.}
  \label{fig5}
\end{center}
\end{figure}

When $\Delta$ is turned on, this BIC becomes a quasi BIC, Now, this state contributes with an extra Fano antiresonance to the conductance. Besides, the conductance displays four resonances while tdue to the breaking of the degeneracy of the resonances. For instance, for $\Delta=0.1\Gamma$ a sharp antiresonance appears at $\varepsilon_f=0$ due to the quasi-BIC and the resonance around $\varepsilon_f=0$ splits in two resonances. This split  grows as $\Delta$ increases ($\Delta=\Gamma$) as we can see in Fig.\ref{fig3}. The separation between resonances grows and the width of the lateral Fano line shapes decreases with $\Delta$. In fact, Fig.\ref{fig5} displays as $\nu_{\alpha}^2$ evolves with $\Delta$. On one hand, $\nu_1^{2}$ (solid black line) vanishes at $\Delta=0$ and grows as $\Delta$ increases and saturates for $\Delta \gg \Gamma$. On the other hand, $\nu_{2,3}^{2}$ decreases almost quadratically with $\Delta$ and tends to zero for the unphysical limit $\Delta\rightarrow\infty$. In this limit, two of the molecular states become BICs. However we can see from Figs.\ref{fig3} and \ref{fig5}, for approximately that for $\Delta \sim 3\Gamma$ these two states become quasi BICs.

\begin{figure}[h]
\begin{center}
 \includegraphics[width=7cm]{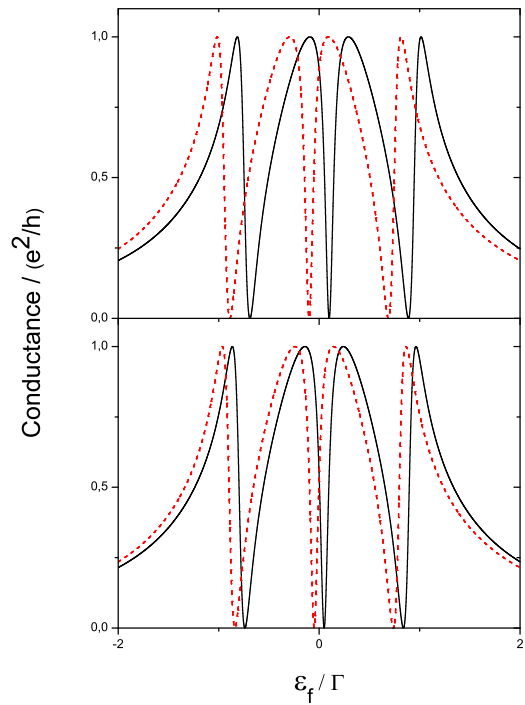}\\
  \caption{(Color online) Spin dependent conductance as a function of the Fermi energy, $\sigma=\,\downarrow$ (dashed red line) and for $\sigma=\,\uparrow$ (solid black line). The upper panel corresponds to $\varepsilon_{z}=0.1\,\Gamma$ and the lower panel to $\varepsilon_{z}=0.05\,\Gamma$. Here we have same considerations than Fig.~(\ref{fig3}), but $\Delta=0.35\,\Gamma$.}
  \label{fig6}
\end{center}
\end{figure}

\begin{figure}[h]
\begin{center}
 \includegraphics[width=8cm]{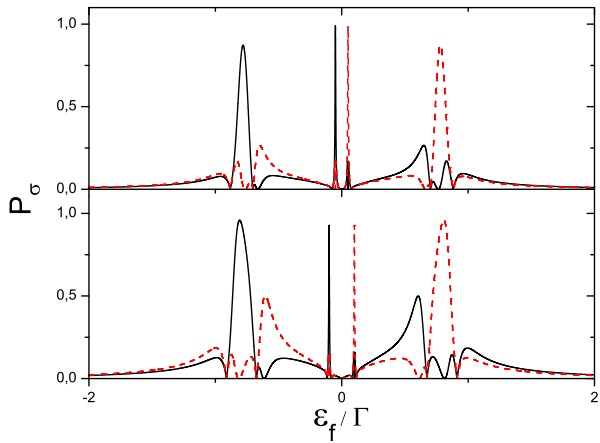}\\
  \caption{(Color online) Weighted spin-polarization for $\sigma=\,\downarrow$ (dashed red line) and for $\sigma=\,\uparrow$ (solid black line), for $\Delta=0.1\,\Gamma$ and Zeeman energy $\varepsilon_{z}=0.1\,\Gamma$ (lower panel)and $\varepsilon_{z}=0.05\,\Gamma$ (upper panel).}
  \label{fig7}
\end{center}
\end{figure}

In presence of a magnetic field, the spin degeneracy is broken and the conductance becomes spin dependent. Fig.\ref{fig6} displays the spin dependent conductance as a function of the Fermi energy for two values of magnetic field, $\varepsilon_{z}=0.1\Gamma$ (upper panel) and $\varepsilon_{z}=0.05\Gamma$ (lower panel) with $\Delta=0.2\Gamma$. By adjusting $\Delta$ and magnetic field, it is possible to control the position and width of the resonances and antiresonances in the conductance. This property can be used in order to produce a high spin-polarization, turning a resonance for one spin with the antiresonance for the opposite spin. In fact, Fig. \ref{fig7} displays the spin-polarization for two values of the magnetic field, $\varepsilon_z=0.1\Gamma$ (lower panel) and $\varepsilon_z=0.05\Gamma$ (upper panel) for fixed $\Delta=0.1\Gamma$. We can observe that the system shows optimal spin-polarization in a wide energy range.

\begin{figure}[ht]
\begin{center}
\includegraphics[width=7cm]{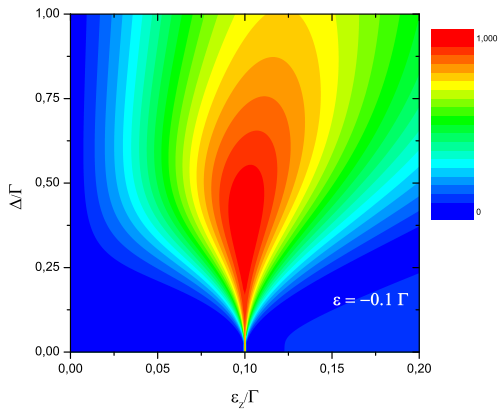}\\
  \caption{(Color online) Weighted polarization at the energy $\varepsilon_f=-0.1\,\Gamma$ in the space of the parameters $\Delta$ and $B$}\label{fig8}
\end{center}
\end{figure}

In order to show the robustness of the high spin-polarization under changes of the parameters of the system, Fig.\ref{fig8} displays a contour plot with the weighted spin-polarization in the space of the parameters $\Delta$ and Zeeman energy $\varepsilon_{z}$ for a  fixed Fermi energy ($\varepsilon_{f}=-0.1\Gamma$). This demonstrates that the spin-polarization filtering capabilities of our device can be made optimum ($\sim 100\%$) within a wide range of parameters.

\section{Summary}
In summary, we have studied the formation of bound states in the continuum  in a side attached quantum dot molecule dot. We show that the bound states in the continuum are formed by the internal interference in the quantum dot molecule. Besides we propose a novel spin filter device based on the combination of bound states in the continuum and Fano effect. Assuming perfectly coherent transport, we have demonstrated that the spin-polarization filtering capabilities of our device can be made optimum ($\sim 100\%$) within a wide range of parameters (Fermi energy and applied magnetic field). It is important to emphasize that the proposed system can exhibit high spin polarization even for very weak magnetic fields.

%Candidate systems to carry out the proposed spin filter are quantum dots build by electrostatic means [13, 14] in a two- dimensional %electron gas in a semiconductor heterostructure of AlGaAs/GaAs, and also graphene quantum dots [1517].

\section{Appendix}\label{ape}

In this section, we show the re-normalization process used to obtain the effective model shown in Fig.\ref{fig1} (lower panel).

% Is easy to obtain a transmission probability for a QD embedded in a continuum using the Green's functions method, because %the Green's function of an isolated QD of energy $\varepsilon_{d}$ is $1/(\varepsilon-\varepsilon_{d})$, and the %corresponding transmission probability for the system considering both contacts equals, by using the Fisher-Lee relation, is
%\
%begin{equation}\label{11}
%\tau=\frac{\Gamma^{2}}{(\varepsilon-\varepsilon_{d})^{2}+\Gamma^{2}}\,.
%\end{equation}
%\\

%Our idea is to use the equations above for a effective QD which consider the contribution of the complete artificial molecule.

By using the Dyson's equations for the Green's functions we can obtain a set of equation for the model indicated in Fig. \ref{fig1}.  In first place, by decimating the external quantum dots (2 and 3) into the central quantum dot (1), we obtain the following equations for the Green's functions,
\begin{subequations}
\begin{eqnarray}\label{a1}
G_{11}&=&g_{1}+g_{1}\,t\,G_{21}+g_{1}\,t\,G_{31}\\
G_{21}&=&g_{2}\,t\,G_{11}\\
G_{31}&=&g_{3}\,t\,G_{11}\,,
\end{eqnarray}
\end{subequations}
so we have $G_{11}$
\begin{equation}\label{a2}
G_{11}=\frac{1}{\varepsilon-\varepsilon_{1}-\frac{t^2}{\varepsilon-\varepsilon_{2}}-\frac{t^2}{\varepsilon-\varepsilon_{3}}}\equiv\tilde{g}_{1}\,,
\end{equation}
we note that
\begin{equation}\label{a3}
\tilde{\varepsilon}_{1}\equiv\varepsilon_{1}-t^2[(\varepsilon-\varepsilon_{2})^{-1}+(\varepsilon-\varepsilon_{3})^{-1}]\,,
\end{equation}
and
\begin{equation}\label{a4}
\Sigma_{\text{m}}(\varepsilon)=\frac{\nu^{2}}{\varepsilon-\tilde{\varepsilon}_{1}}\,.
\end{equation}

Now we decimate the normalized lateral part with the embedded QD in a continuum (QD0), so we have
\begin{subequations}
\begin{eqnarray}\label{a5}
G_{00}&=&g_{0}+g_{0}\,\nu\,\tilde{G}_{10}\\
\tilde{G}_{10}&=&\tilde{g}_{1}\,\nu\,G_{00}\,.
\end{eqnarray}
\end{subequations}

From above equations and considering the Zeeman's effect, we obtain the Eq.\ref{eq6}.

\section*{Acknowledgments}\label{agradecimientos}
Authors thank the funding of the FONDECYT project number 1100560 and the scholarship CONICYT academic year 2012 number 22121816.

\end{document}